\documentclass{emulateapj}
\usepackage{epstopdf}

\usepackage{natbib}

\newcommand{\ms}{M$_{\odot}$}
\newcommand{\ls}{L$_{\odot}$}
\newcommand{\kms}{$\,\rm km\,s^{-1}$}

\newcommand{\hh}{H$_2$}

\newcommand{\eqq}{\!=\!}  
\newcommand{\too}{\!\!\rightarrow\!\!} 
\newcommand{\jone}{{$J\eqq 1\too0$}}
\newcommand{\jtwo}{{$J\eqq2\too1$}}

\newcommand{\jsix}{{$J\eqq6\too5$}}

\newcommand{\jeight}{{$J\eqq8\too7$}}
\newcommand{\jnine}{{$J\eqq9\too8$}}
\newcommand{\jten}{{$J\eqq10\too9$}}
\newcommand{\jeleven}{{$J\eqq11\too10$}}
\newcommand{\jtwelve}{{$J\eqq12\too11$}}
\newcommand{\jthirteen}{{$J\eqq13\too12$}}
\newcommand{\wone}{$\rm 2_{0,2}\too1_{1,1}$}
\newcommand{\wtwo}{$\rm 3_{1,2}\too3_{0,3}$}
\newcommand{\wthree}{$\rm 1_{1,1}\too 0_{0,0}$}
\newcommand{\wfour}{$\rm 3_{1,2}\too 2_{2,1}$}
\newcommand{\wfive}{$\rm 3_{2,1}\too 3_{1,2}$}
\newcommand{\wsix}{$\rm 4_{2,2}\too 4_{1,3}$}
\newcommand{\wseven}{$\rm 2_{2,0}\too 2_{1,1}$}
\newcommand{\weight}{$\rm 5_{2,3}\too 5_{1,4}$}
\newcommand{\fdeg}{^\circ}
\newcommand{\as}{$^{\prime\prime}$}
\newcommand{\asm}{^{\prime\prime}}
\newcommand{\mm}{$\rm \mu m$}
\newcommand{\by}{\!\times\!}
\newcommand{\water}{$\rm{H_2O}$}

\newcommand{\apm}{APM~08279+5255}

\shorttitle{APM 08279+5255 Water Spectrum}
\shortauthors{Bradford et al.}
\begin{document}
\bibliographystyle{apj}  
\title{The Water Vapor Spectrum of APM~08279+5255:  X-Ray Heating and Infrared Pumping over Hundreds of Parsecs }

\slugcomment{Accepted for Publication in the Astrophysical Journal Letters}

\author{C.M. Bradford\altaffilmark{1,2}, A.D. Bolatto\altaffilmark{3}, P.R. Maloney\altaffilmark{4}, J.E. Aguirre\altaffilmark{4,5}, J.J. Bock\altaffilmark{1,2},  J. Glenn\altaffilmark{4}, J. Kamenetzky\altaffilmark{4}, R. Lupu\altaffilmark{5}, H. Matsuhara\altaffilmark{7}, E.J. Murphy\altaffilmark{7}, B.J. Naylor\altaffilmark{2,1}, H.T. Nguyen\altaffilmark{1}, K. Scott\altaffilmark{5}, J. Zmuidzinas\altaffilmark{2,1}}

\altaffiltext{1}{Jet Propulsion Laboratory, California Institute of Technology, Pasadena, CA, 91109, USA}
\altaffiltext{2}{California Institute of Technology, Pasadena, CA, 91125, USA}
\altaffiltext{3}{University of Maryland, College Park, MD, 20742-2421, USA}
\altaffiltext{4}{University of Colorado, Boulder, CO, 80303, USA}
\altaffiltext{5}{University of Pennsylvania, Philadelphia, PA, 19104, USA}
\altaffiltext{6}{Institute for Space and Astronautical Science, Japan Aerospace and Exploration Agency, Sagamihara, Japan}
\altaffiltext{7}{Observatories of the Carnegie Institution for Science, 813 Santa Barbara Street, Pasadena, CA 91101, USA}

\begin{abstract} 
We present the rest-frame 200--320~\mm\ spectrum of the z=3.91 quasar \apm, obtained with Z-Spec at the Caltech Submillimeter Observatory.  In addition to the \jeight\ to \jthirteen\ CO rotational transitions which dominate the CO cooling, we find six transitions of water originating at energy levels ranging up to 643~K.  Most are first detections at high redshift, and we have confirmed one transition with CARMA.  The CO cooling is well-described by our XDR model, assuming L$_{\rm 1-100\,keV}\sim1\times10^{46}\rm\,erg\,s^{-1}$, and that the gas is distributed over a 550-pc sizescale, per the  now-favored $\mu$=4 lensing model.  The total observed cooling in water corresponds to 6.5$\times$10$^{9}$~\ls, comparable to that of CO.  We compare the water spectrum with that of Mrk~231, finding that the intensity ratios among the high-lying lines are similar, but with a total luminosity scaled up by a factor of $\sim$50.   Using this scaling, we estimate an average water abundance relative to \hh\ of 1.4$\times10^{-7}$, a good match to the prediction of the chemical network in the XDR model.  As with Mrk~231, the high-lying water transitions are excited radiatively via absorption in the rest-frame far-infrared, and we show that the powerful dust continuum in \apm\ is more than sufficient to pump this massive reservoir of warm water vapor.  
\end{abstract}


\section{Introduction}

\subsection{Water as a Molecular Gas Coolant and Radiation Field Probe}
Though virtually impossible to observe in the local Universe from the ground, \water\ is an important constituent of interstellar gas under the right conditions.   The species is chemically enhanced for temperatures above a few hundred Kelvin \citep{Draine_82}, and is sublimated from grain mantles as dust is heated to $\sim$100~K.  In these conditions, it becomes an important gas-phase repository for oxygen, with abundance approaching that of CO (e.g.~\citet{Cernicharo_06a}).   With water's large dipole moment, the network of rotational transitions can either be a dominant molecular coolant (for $T>300\,\rm K$ and $n_{\rm H_2}>10^{5}\rm cm^{-3}$; \citet[and references therein]{Hollenbach_79, Neufeld_87, Neufeld_93}), or a pathway for coupling mid- and far-infrared continuum radiation into the gas if the local continuum is sufficiently strong (e.g. \citet{Scoville_76}. 

Many of the most important \water\ transitions lie in the short submillimeter, between 200--300~\mm, which has until recently been inaccessible in the local universe, and has been only sparsely explored at high redshifts via observations in the millimeter band.   With the advent of broadband spectrometers enabled by high-performance bolometer arrays (Herschel SPIRE--\citet{Griffin_10} and Z-Spec--e.g.~\citet{Earle_06}), the water spectrum can be explored on galaxy scales locally and at high redshift.  Water is not an important coolant in the overall energy budget of the gas in the Milky Way (via COBE, \citet{Fixsen_99}) or even in the nucleus of the nearby starburst galaxy M82 \citep{Panuzzo_10}.   However, in extreme ULIRGs such as Arp~220 \citep{Rangwala_11}, and Mrk~231 (\citet[hereafter vdW10]{vanderWerf_10}, \citet[hereafter GA10]{Gonzalez-Alfonso_10}), the most luminous \water\ transitions lines can be comparable in strength to those of CO.   

Through comparison with water excitation models, GA10 show that the water emission lines in Mrk~231 measured with SPIRE are not produced by purely collisional excitation; the ratio of high-level transitions to the lower-level and ground-state transitions is too large, even for shock-like conditions.
They present a model in which the 200--300~\mm\ water emission spectrum is pumped by the absorption of shorter-wavelength continuum photons, a scenario supported by absorption measurements with the ISO LWS \citep{Gonzalez-Alfonso_08}.   The excitation of water is thus coupled closely to the dust continuum intensity, and GA10 are able to use the ISO + SPIRE lines to constrain the size and opacity of the dust continuum source in Mrk~231 between 30 and 150 microns, concluding that it is $\sim$110--180~pc in size.

We might expect to find similar processes at work in the early Universe, and our wideband 1 mm surveys with Z-Spec are revealing \water\ in a handful of high-redshift objects \citep[Bradford et al.;~2009 (hereafter B09),][]{Lupu_11,Omont_11}.   We present here a study of \apm\ at z=3.91, a particularly convenient redshift for characterizing the water emission spectrum in the 1~mm band. 
 
 \begin{figure*}[t!]\centering
\includegraphics*[width=16cm,bb=10 13 650 440]{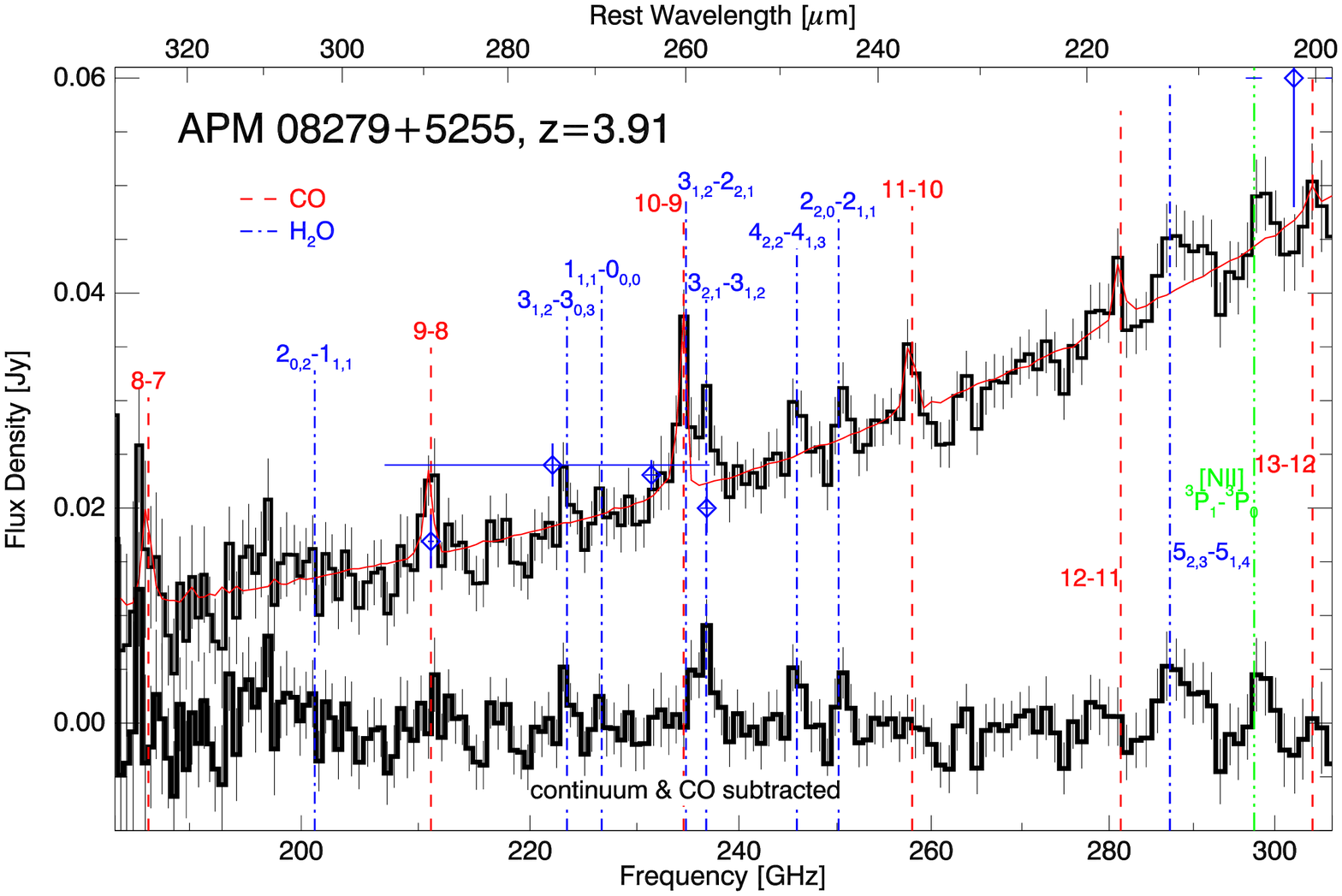}
\caption{Z-Spec spectrum of \apm.   Dashed vertical lines show all transitions of CO and \water\  with upper-level energies below 700~K which lie in the band.  The position of the [NII]~205~\mm\ transition is also noted.   Diamonds show continuum measurements from (in increasing frequency): Plateau de Bure (W07),  SCUBA \citep{Lewis_98}, CARMA (this work), CARMA (this work), and the SMA (average of two sidebands---horizontal and vertical bars indicate bandwidth and uncertainty) \citep{Krips_07}.    The lower spectrum shows the data with a fit to the continuum and CO lines subtracted.}
\label{fig:spec}
\end{figure*}

\subsection{\apm}
\apm\ is an unusual source by any measure. It exhibits an apparent total bolometric luminosity of $7\times10^{15}$~\ls, dominated by warm AGN-heated dust peaking in the rest-frame 15--30~\mm\ regime \citep[and references therein, hereafter R09]{Riechers_09}.  
There is a cold dust component identified in the rest-frame far-IR and submillimeter which contributes $\sim$3\% of the total energy.  This may be produced by star formation, and the [OIII] 88~\mm\ luminosity (1.06$\times$10$^{11}$/$\mu$~\ls) is consistent with star formation contributing up to $\sim$35\% of this far-IR luminosity \citep{Ferkinhoff_10}, or $\sim$1\% of the total in the system.  \apm\ has been studied in detail with interferometric observations of CO ranging from \jone\ and \jtwo\ with the VLA (\citet{Papadopoulos_01}, more recently R09) to \jeleven\ with the Plateau de Bure Interferometer (PdB) \citep[hereafter W07]{Weiss_07}.   The compact CO morphology (at least at \jone) is similar to the image of the optical quasar---only just marginally resolved into sub-images at 0.4\arcsec, and R09 model the CO \jone-emitting gas as a R$\sim$550~pc disk, magnified by $\mu$=4.   We adopt this magnification factor, and a luminosity distance of 35,565~Mpc (appropriate for $H_0=71$, $\Omega_M=0.27$, $\Omega_\Lambda=0.73$).   The W07 CO measurements then translate to an intrinsic molecular gas mass of $\rm M_{total}\sim1.0\times10^{11}$~\ms\ using an adopted scale factor of $5.3\times10^{11}\,\rm M_\odot\,(K\,km\,s^{-1}\,pc^2)^{-1}$.   

In the R09 lensing model, the magnification is not sharply peaked in the central few hundred parsecs:  $\mu(R)$ ranges from 3--5 out to source radius $R\sim$10~kpc.  If this is correct, then the CO line ratios and the gas conditions inferred from CO (W07) are largely independent of the lensing, and the high-J CO emission in \apm\ is particularly unusual among extragalactic sources.   To reproduce the ratios among the low-J and high-J CO transitions, 
W07 require at least two components with conditions ranging from cool and dense (T$\sim$65~K$, \rm n_{H_2}\!\!\sim\!\!1\by 10^5\,cm^{-3}$) to warmer and less dense (T$\sim$225~K,  $\rm n_{H_2}\!\!\sim\!\!1\by 10^4\,cm^{-3}$).  

\section{Observations}\label{sec:obs}

\apm\ was observed with Z-Spec at the Caltech Submillimeter Observatory (CSO) on a total of 13  nights among 4 observing campaigns:  
2008 March 23--28, 2008 April 4--16, 2009 January 6--9 and 2009 February 23--27.  We used the chop-and-node mode with a 20 second nod period and a 1.6~Hz, 90\as\ chop.  The total integration time (including chopping) is 25.3 hours, with a range of telluric water vapor burdens; $\tau_{225\rm GHz}$ ranged from 0.04 to 0.20, with a median of 0.083 and a mean of 0.102.   With all data incorporated in a weighted coaddition, the sensitivity referred to the total integration time ranges from 750~mJy~s$^{1/2}$ in the middle of the band (producing a 2.5 mJy channel RMS) to 1100~mJy~sec$^{1/2}$ at the band edges (3.6~mJy channel RMS), with greater degradation below 200~GHz.  The data are calibrated using Mars and Uranus, with quasars as secondary calibrators using the method described in B09.  Figure~\ref{fig:spec} shows the spectrum and the line fits described below.

\subsection{CARMA Follow-Up}
\begin{figure}[t!]\centering
\includegraphics*[width=8.8cm]{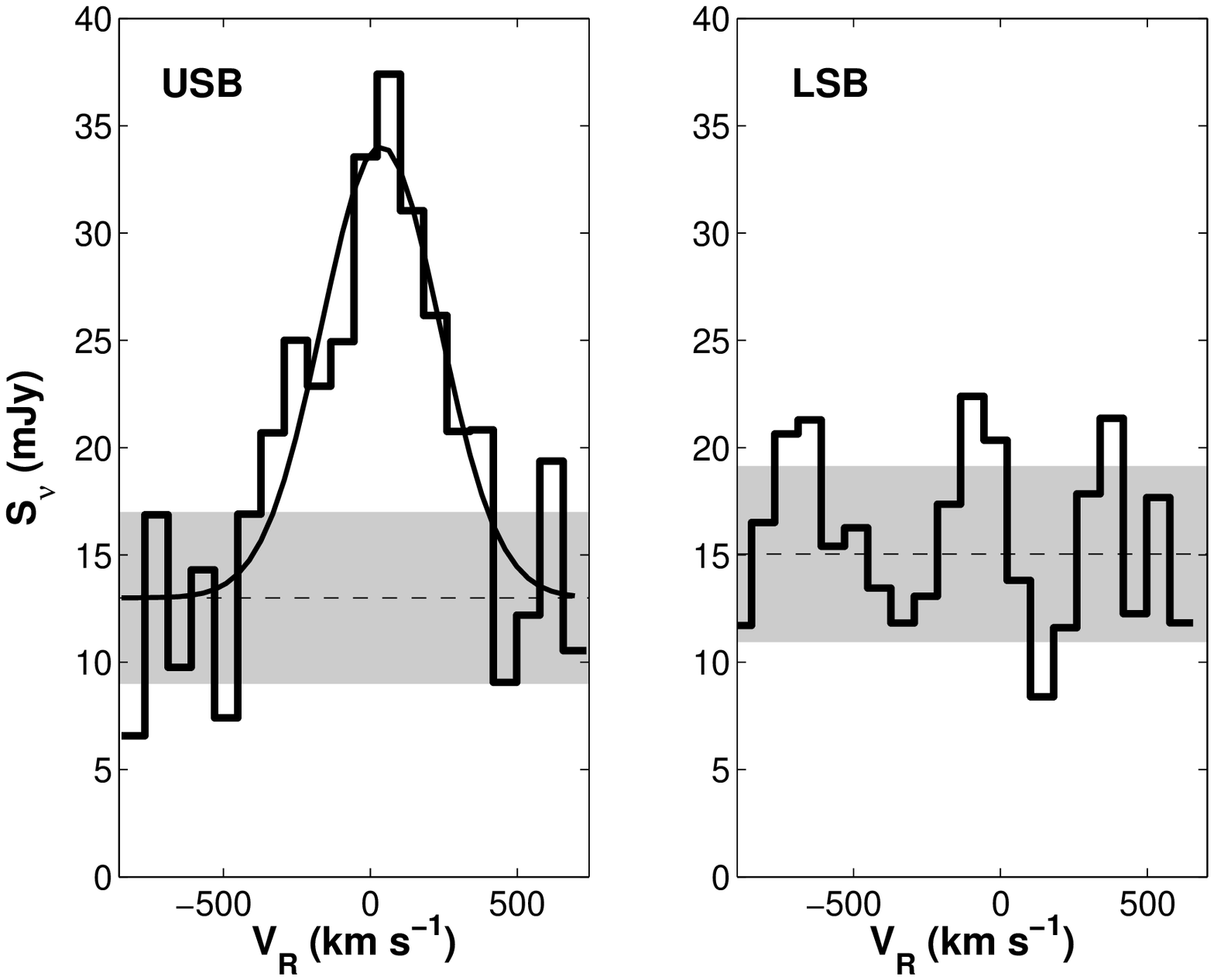}
\caption{CARMA spectrum of the \wfive\ \water\ transition.   The velocity scales refers to $z=3.9119$.   The plotted spectra are not corrected for the atmospheric decorrelation, though the entries in Table~1 are.\label{fig:carma}}
\end{figure}

The source was also observed using the 15-element Combined Array for
Research in Millimeter-wave Astronomy 
(CARMA, \citep{Bock_06}),  (Figure~\ref{fig:carma}).
On 2009 November 9 and 15, with
the array in its C configuration.  The 
upper sideband was tuned to $\nu_{USB}=236.75$ GHz, corresponding to the \water\ \wfive\  transition at $z=3.9119$, the redshift measured with PdB for CO \jnine\ (the highest-J transition measured with the PdB) (W07).  The lower sideband was centered on $\nu_{LSB}=231.14$ GHz.  The
correlator was configured in its $3\times468.75$ MHz setting with
partially overlapping windows, yielding an effective coverage of
$\Delta\nu\approx1.3$ GHz ($\Delta v\approx1600$ km~s$^{-1}$) with a
native resolution of $\approx 31.25$~MHz ($\approx 39.6$ km~s$^{-1}$).  

The double sideband system temperature was
T$_{sys}\sim280-480$~K, and the total integration time was
 9 hours.  The quasar $0359+509$ was
the astronomical bandpass calibrator ($S_\nu\sim3.5$ Jy), and Mars was the absolute flux calibrator.  The quasars $0927+390$
($S_\nu\sim2.3$ Jy) and $0920+446$ ($S_\nu\sim1.0$ Jy) were primary and
secondary phase calibrators, respectively, on a calibration cycle of 10
minutes. The phase RMS on $0927+390$ after self calibration was
approximately 25$^\circ$. 
Atmospheric decorrelation effects were estimated 
 to be 35\%, and the fluxes have been corrected accordingly.  This is not
unexpected since the observations were obtained in a mid-size
configuration, with typical baselines of $\sim360$~m.   

The synthesized beam size is
$\theta=0.80\asm\times0.64\asm$ with $PA=-70\fdeg4$.
Fitting to the LSB and the USB images are very 
consistent and suggest that \apm\ is slightly resolved with
an intrinsic angular size of $\theta\approx0.4\asm$, in agreement with other observations of the source angular size~\citep{Krips_07}.

\begin{deluxetable*}{lrcclll}
\tabletypesize{\scriptsize}
\tablecaption{1-mm Line Intensity Measurements in \apm}
\tablewidth{0pt}
\tablehead{
\colhead{transition}&\colhead{E$_{\rm upper}$ [K]}&\colhead{$\nu_{\rm rest} {\rm[GHz]}$}&\colhead{$\nu_{\rm obs}$ [GHz]}&\colhead{flux [Jy km/s]}&\colhead{1$\sigma$ err [Jy km/s]\tablenotemark{a}}&\colhead{L$_{\rm intrinsic}$ [10$^8$ \ls]\tablenotemark{b}}
}
\startdata

CO \jeight &199 & 921.80&  187.7 & 12.1 & 8.3 & 7.4 $\pm$ 5.1\\
CO \jnine & 249 &1036.91 &  211.1 & 16.0 & 3.8 & 	11.0 $\pm$ 2.6\\
\multicolumn{4}{l}{\hspace{0.15in}\it IRAM 30-m (W07)} &\ \ \ \ 11.1 &\ \ \ \ 2.2 &\ \ \ \ \ 7.7 $\pm$ 1.5 \\
\multicolumn{4}{l}{\hspace{0.15in}\it IRAM Plateau de Bure (W07)} &\ \ \ \ 12.5 &\ \ \ \ 2.4 &\ \ \ \ \  8.6 $\pm$ 1.7\\
CO \jten & 304 &1151.99 &  234.6 & 15.2\tablenotemark{c} & 7.2 & 11.6 $\pm$ 5.5 \\
\multicolumn{4}{l}{\hspace{0.15in}\it IRAM 30-m (W07)} &\ \ \ \ 11.9 &\ \ \ \ 2.4 &\ \ \ \ \ 9.1 $\pm$ 1.8\\
CO \jeleven & 365 & 1267.02 &  258.0 & 14.4 & 4.5 & 12.1 $\pm$ 3.8\\
\multicolumn{4}{l}{\hspace{0.15in}\it IRAM 30-m (W07)} &   \ \ \ \  10.4 & \ \ \ \ 2.1 &\ \ \ \ \ 9.2 $\pm$ 1.8 \\
CO \jtwelve &  431 & 1381.00 &  281.4 & 9.5 & 5.0 & 8.8 $\pm$ 4.6\\
CO \jthirteen & 503 & 1496.92 &  304.8 & 4.4 & 6.1 &  \\
\\
\water\ \wone &101 & 987.93 &  201.2 & 1.9 & 4.1 & $<$8.1 (3$\sigma$) \\
\water\ \wtwo & 250 &1097.40 &  223.5 & 8.1 & 3.5 & 5.9 $\pm$ 2.6 \\
\water\ \wthree &  53.4 & 1113.34 & 226.7 & 4.6 & 3.7 & 3.4 $\pm$ 2.7 ($3\sigma\!<\!8.2$)\\
\water\ \wfour & 250 &1153.13 &  234.8 & 14.5\tablenotemark{c} & 7.2 & 11.1 $\pm$ 5.5\\
\water\ \wfive &305 & 1162.91 &  236.8 & 16.0 & 3.5 & 12.4 $\pm$ 3.0\\
\multicolumn{4}{l}{\hspace{0.15in}\it CARMA (this work)} &\ \ \ \ 16.3 &\ \ \ \ 3.5 &\ \ \ \ \ \ 12.6 $\pm$ 2.7 \\
\water\ \wsix & 455 &1207.64 &  245.9 & 12.8 & 4.3 & 10.3 $\pm$ 4.1\\
\water\ \wseven & 196 &1228.79 &  250.2 & 7.9 & 4.4 & 6.4 $\pm$ 4.0 ($3\sigma\!<\!12.0$)\\
\water\ \weight & 643 &1410.62 &  287.2 & 16.0 & 6.3 & 15.0 $\pm$ 5.2\\
\\
NII $^3\!P_1 \rightarrow ^3\!\!P_0$ &70.1 & 1461.0 &  297.4 & 7.6 & 5.4 & 7.4 $\pm$ 5.1\\
\enddata
\tablenotetext{a}{Z-Spec fluxes include statistical uncertainties only, IRAM uncertainties are $\pm$20\%, per W07.}
\tablenotetext{b}{All luminosities are corrected for lensing, assuming a magnification factor of 4.}  
\tablenotetext{c}{CO \jten\ and \water\ \wfour\ are blended, see discussion in text.}
\label{tab:lines}
\end{deluxetable*}

\section{Results, Continuum and Line Flux Extractions}\label{sec:results}

A model spectrum consisting of a a power-law continuum with 15 Gaussian spectral lines is fit to the 160 bolometer flux values, incorporating their measured spectral response profiles using the method described in \citet{Naylor_10}.   The redshift is fixed at 3.9119, and the line width is $\Delta v_{\rm FWHM} = 550 \,\rm km\,s^{-1}$.  This width is greater than that reported by W07 for the low-J transitions, but is consistent with values measured for the \jnine\, \jten\ CO transitions.   In any case, our fits to the Z-Spec spectrum are not very sensitive to the adopted line width.   The fitted fluxes and resulting uncertainties are presented in Table~\ref{tab:lines}.   

Our CO fluxes are consistent with the W07 IRAM measurements to within the uncertainties, though we do find systematically larger values, e.g. for \jnine\ and \jeleven\ (35--45\%).     Since the continuum measurements are consistent to better than this (see Figure~\ref{fig:spec}) we rule out global calibration problems at this level.  It may be that Z-Spec is coupling flux in high-velocity wings of the lines which would be missed in the IRAM spectra, or there could be spatially extended CO emission that the interferometer resolves out.   Given the coarse resolution, it is also possible that other transitions near CO are contaminating the Z-Spec fitted flux, though aside from the \jten\ transition, there are no significant candidates in the SPIRE Mrk~231 spectrum \citep{vanderWerf_10}.   In any case, our CARMA measurement is consistent with Z-Spec in both the line flux and continuum to better than the statistical uncertainties.

CO \jten\ is clearly blended with the \water\ \wfour\ transition; the two transitions are separated by 297~\kms, about 1/3 of a Z-Spec channel.  We fit the two simultaneously, but only their sum is meaningful.   We estimate the flux of CO \jten\ as the geometric mean of the \jnine\ and \jeleven\ intensities in temperature units, and then subtract this value from the fitted flux of the CO+water blend to arrive at an estimate of the flux for the water line.   We find that the water and CO fluxes are comparable, though the formal SNR for each is modest ($\sim2.0$) as the errors are propagated in this estimate.

Figure~\ref{fig:spec} and Table~\ref{tab:lines} show all eight \water\ transitions in the band with upper level energies less than 700~K.   Of these, we detect four in emission with $>$2.3$\sigma$ significance.  While less than the customary 3$\sigma$, 2.3$\sigma$ corresponds to a likelihood of only $(1-e\!r\!f({S\!N\!R/\sqrt{2})})/2$ = 1.0\% that the transition is not real with positive flux, if the noise is Gaussian.  In addition to \wfour\ described above (2$\sigma$ is only 2.2\% likely to be false), we tentatively identify two additional transitions in emission at lower significance (\wthree\ and \wseven) and estimate fluxes and upper limits.   \wone\ can only offer an upper limit.   In Figure~3, we plot 2$\sigma$ upper limits, corresponding to a 98\% confidence level.

The ground-state fine-structure transition of [NII] ($^3\!P_1 \rightarrow ^3\!\!P_0$, $\lambda_{rest}$=205.18~\mm) is identified as tentatively in emission at low significance (1.4$\sigma$).   The value is consistent with the lower limit obtained by \citet{Krips_07}  with the SMA (3$\sigma<9\rm\,Jy\,km\,s^{-1}$ in a single beam beam, 3$\sigma<16\rm\,Jy\,km\,s^{-1}$ if scaled to the continuum source size \citep{Ferkinhoff_10}).  

\section{Analysis and Discussion}

\subsection{CO Emission and XDR heating}
We find that the ratios among the high-J CO transitions are consistent with the excitation model presented in W07.   If the IRAM fluxes are increased to match the Z-Spec values with the factor of 1.4 described above, our \jeight\ and \jtwelve\ fluxes and the \jthirteen\ limit all fall on the CO excitation curves in Figures~7 and 9 of W07, matching to well within the statistical uncertainties.   We estimate a total lensing-corrected CO luminosity of $\sim$7$\times$10$^{9}$~\ls, a fraction 1.4$\times$10$^{-4}$ of the 5$\times$10$^{13}$~\ls\ output in the far-IR per the decomposition of R09.   This total CO luminosity fraction is not  unusually high; it is comparable to the value in local starburst galaxies and ULIRGs for which the spectra have now been measured with Herschel SPIRE.   $\rm L_{CO} / L_{IR}$ is 1.25$\times$10$^{-4}$ in M82 \citep{Panuzzo_10},  1.0$\times$10$^{-4}$ in Arp~220 \citep{Rangwala_11}, and 7$\times$10$^{-5}$ in Mrk~231 (vdW10).   However, as W07 have noted, the CO excitation of \apm\ is much higher:  the \jeleven\ to \jsix\ luminosity ratio, for example ranges from $0.54\pm0.13$ in M82 to $\sim1$ in Mrk~231 to 2.5--3.5 in \apm, depending on the calibration.    Since the high-lying CO transitions in Mrk~231 cannot be understood in a reasonable PDR framework, the even more extreme ratios in \apm\ cannot be fit by PDR models either.   

Given the powerful hard X-ray source, it is likely the high-J CO emission is due to gas cooling in X-ray dominated regions (XDRs, \cite{Maloney_96}), as has been modeled for both the Mrk~231 and the Cloverleaf system (B09).   For \apm, we estimate an intrinsic 1--100~keV luminosity of 1$\times$10$^{46}\,\rm erg\,s^{-1}$, by extrapolating the apparent luminosity in the rest-frame 2--10~keV range of 2.3$\times$10$^{46}\,\rm erg\,s^{-1}$ per \citet{Just_07}.   This is similar to the hard X-ray luminosity inferred for the Cloverleaf, and if we assume the same attenuating column of $\rm N_{H,att}=3\times10^{23}\,cm^{-3}$ (as is required to maintain a molecular phase), then we can use the same model as in B09 to consider the CO excitation (Figures 6, 7, 8 in B09).  Assuming the CO emission emerges from a 550-pc disk as per R09, then the CO surface brightness ($\rm erg\,s^{-1}\,cm^{-2}$) in \apm\ is 0.7--0.8~$\rm erg\,s^{-1}\,cm^{-2}$, some 3.5--4 times that of the Cloverleaf.   This emergent flux is readily produced with R$<$650~pc and n$_{\rm H}$$>$2$\times$10$^5\rm\,cm^{-3}$ (n$_{\rm H_2}$$>$1$\times$10$^5\rm\,cm^{-3}$ if fully molecular).  While the required density is on the high end of that inferred in the W07 CO analysis; this is overall a good match to the observed parameters.

\subsection{Water Emission and Radiative Pumping.}

Totaling the available transitions in our band, we find a total luminosity of 6.5$\times$10$^9$~\ls\ in the water rotational network.  This is a lower bound, but is already close to the total CO emission, suggesting that water is energetically important to the gas.
As GA10 found for Mrk~231, the ratio of luminosity in the high-excitation transitions relative to the low-lying ones indicates that the high-excitation levels are not excited collisionally.   The conditions inferred by W07 using CO (temperature from 70--200~K and n$_{\rm H_2}$ between 10$^4$ and 10$^5\,\rm cm^{-3}$) are well below even the extreme conditions considered by GA10 (T=200~K, n$_{\rm H_2}$=1.5$\times$10$^6\,\rm cm^{-3}$) in considering collisional excitation.   

\begin{figure}[t!]\centering
\includegraphics*[width=8.8cm,bb=25 5 450 445]{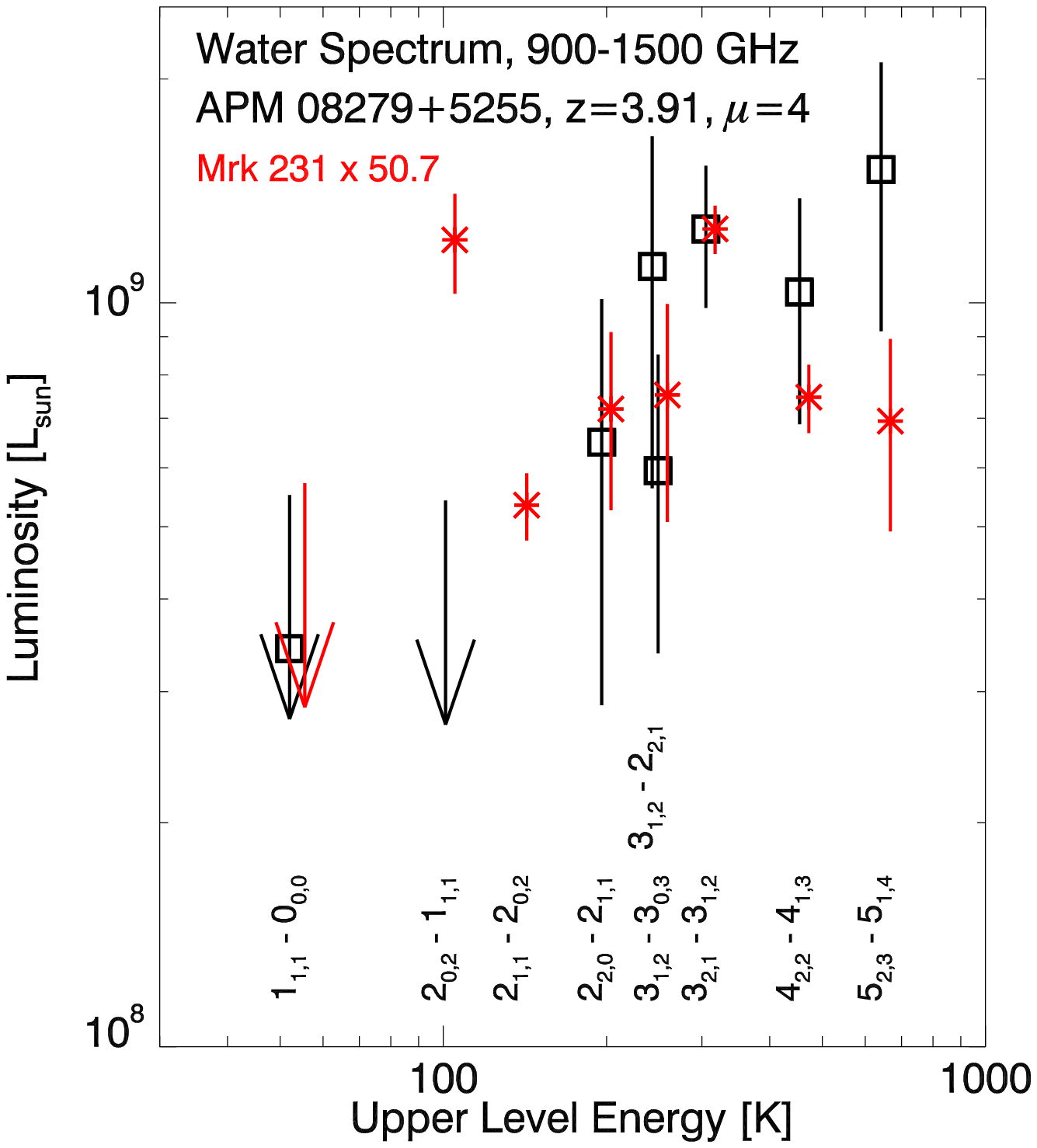}
\caption{Comparison of the water spectrum in Mrk~231 with that in \apm, as measured with Z-Spec.   Downward arrows show limits at 2$\sigma$.  The \water\ \wfour\ flux is not measured in Mrk~231, though as vdW10 note, the \jten\ transition likely includes flux from this line.\label{fig:compare}}
\end{figure}

Figure~\ref{fig:compare} shows a comparison of the \apm\ \water\ spectrum to that of Mrk~231 (adopting d$_{\rm L}$=184~Mpc).  To our measurement accuracy, the emission in the high-lying lines in \apm\ resembles that in Mrk~231, scaled up by $\sim$50$\times$.  In fact, \apm\ shows an even higher ratio of the E$_{\rm upper}>200~\rm K$ transitions to the lower-energy ones.   There is plenty of molecular material in the source---the R09 molecular gas mass estimate is a factor of 22 times the GA10 estimate for Mrk~231, including all 3 components, but it is some 300$\times$ the modeled warm component which dominates the high-lying water emission in Mrk~231.  Thus even modest water abundance throughout, or an abundance similar to that obtained in Mrk~231 in 1/6 of the total gas reservoir would be sufficient to produce the observed luminosity in \apm, if it is suitably excited.

The low significance of the detections and lack of higher-frequency coverage for the absorption transitions does not warrant a detailed radiative transfer model, but it is  easy to show that the powerful dust continuum in \apm\ can readily pump the observed water transitions with the same mechanism as that modeled by GA10 for Mrk~231.    Assuming spherical symmetry, the observed flux constrains the size of a fiducial region $R_s$ around a dust emitting source via:
\begin{equation}
\left( \nu F_\nu \right) _{obs}\,\mu^{-1} = \frac{\left(\nu L_\nu\right)_{em}}{4\pi d^2_L} = \frac{R^2_{s}}{d_L^2}\, \nu_{em}\int{I_{\nu\,em}\, d\Omega} 
\end{equation}
where $d_L$ is the luminosity distance, and $\mu$ is the magnification factor.   $I_\nu(\Omega)$ is the specific intensity produced by the dust distribution; it captures brightness temperature and opacity as a function of viewing angle.   For a pumping transition at $\nu_{em}$ this integral of $I_\nu$ over solid angle is the quantity that sets the pumping rate per molecule in a given transition.  Holding this constant allows an estimate of the ratio of the size scale over which similar pump conditions exist in \apm\ as those in Mrk~231, through the scaling:
\begin{equation}
R_{s} \propto d_L \left[\left(\nu F_\nu\right)_{obs}\mu^{-1}\right]^{1/2},
\end{equation}
giving
\begin{equation}
V_{w} \propto R_s^3 \propto d_L^3\left[\left(\nu F_\nu\right)_{obs}\mu^{-1}\right]^{3/2}
\end{equation}
As an example we consider pumping of the 3$_{1,3}$ level with absorption at 58\mm, producing the \wsix\  transition in emission.  Given the observed fluxes, the scaling above suggests that R$_{s}$ is 5.1 times larger for \apm\ than for Mrk~231.   For similar geometries, then, the continuum-emitting source in \apm\ is capable of illuminating a volume $\sim R_s^3 \sim 130$ times larger than Mrk~231 with a similar pump rate.  This is a reasonable match to the factor of 50--90 larger luminosity in this transition in \apm.   We expect that for \apm, the higher color temperature would preferentially pump via higher-frequency far-IR transitions relative to Mrk~231, and this likely explains the increased flux in the \weight\ transition (pumped via 45~\mm) and the reduced flux in \wfour\ (pumped via 75~\mm).

The Mrk~231 warm water-emitting component is modeled by GA10 as a sphere of dust and gas of size 110--180~pc.  With the above scaling, we know that  that similar pump conditions must exist in \apm\ out to distances of 560--920~pc.   This may seem extreme, but \apm\ is an extremely powerful system; consider that the smallest possible physical size for the 220~K source is in the limit of an optically-thick (at $\lambda_{\rm peak}$) sphere, for which R$_{min}$$\sim$215~pc.  Thus if the molecular gas is really situated in a R$\sim$550-pc disk per the model of R09, then a large fraction of the molecular mass in this system is capable of being pumped to the level of the warm component in the Mrk~231 model.   Indeed, the simple model suggests that the water must exist at such large distances because the system is so opaque.   If we require that the same 220~K, 215-pc dust component account for the emission at $\lambda_{\rm rest}$=250~\mm, then it must have an opacity of $\sim$0.8 at this wavelength.  Any contribution from lower-temperature material requires a combination of larger opacity and/or larger physical size.  In the spherically-symmetric model then, this indicates either that the observed fluxes in the water lines must be extinction corrected with large factors, or more likely, that most of the observed line emission originates at R$\ge$200~pc.   This interpretation is consistent with that of \citet{Srianand_00} who find that radiative excitation from a far-IR source larger than the UV source, and at least 200~pc, is required to explain the presence of [SiII] and [CII] in the upper far-IR levels indicated in optical absorption studies.

Without more detailed geometric constraints, we can only conclude that the total mass of water vapor is at least $\sim$50 times that of Mrk~231, some $2.5\by10^4$~\ms, and is likely distributed over scales larger than 200~pc, comparable to the 550-pc size-scale inferred from the CO imaging.   The mass of water vapor corresponds to an average abundance relative to \hh\ of $\ge1.4\times10^{-7}$, when referred to the full molecular gas mass in \apm.  This estimate is a good match to our our XDR model, which predicts a water abundance of 1.4--2.0$\times10^{-7}$ with the parameters described above.  

\acknowledgements{We are indebted to the staff of the Caltech Submillimeter Observatory for their help in Z-Spec's commissioning and observing.  We acknowledge the following grants and fellowships:  NASA SARA grants NAGS-11911 and NAGS-12788, NSF AST grant 0807990, an NSF Career grant (AST-0239270) and a Research Corporation Award (RI0928) to J. Glenn, a Caltech Millikan and JPL Director's fellowships to C.M.B., a NRAO Jansky fellowship to J. Aguirre,  NASA GSRP fellowship to L. Earle, and an NSF GSRP award to J. Kamenetzky.  The research described in this paper carried out at the Jet Propulsion Laboratory, California Institute of Technology, was done so under a contract with the National Aeronautics and Space Administration.   

Copyright 2011. All rights reserved.}


\end{document}